Comment on "High Resolution Polar Kerr Effect Measurements of $Sr_2RuO_4$:
Evidence for Broken Time-Reversal Symmetry in the Superconducting State"

The symmetry of the order parameter plays a crucial role for understanding the superconductivity. Recently, Xia el al. [1] reported the observation of the Kerr angle in large domains, claimed to show in an "unambiguous way" that $Sr_2RuO_4$ breaks time reversal symmetry below Tc, and the p-wave type order parameter vector **d** = **z**[$k_x$ ± i $k_y$] variety in a superconducting $Sr_2RuO_4$ . Unfortunately, their estimation of the Kerr angle of prad, using the formula based on the p-wave order parameter, failed to account for the data. And assuming ambiguously the asymmetry of electron-hole at the Fermi level, they estimated the Kerr angle of 100 (actually 12.5 with the proper factor of index of refraction) nrad compared to the data of 65 nrad. Neither the data supporting for the vector nature of the order parameter, nor the quantitative analysis of the Kerr angle was given. Thus, their claim, in an "unambiguous way", of the p-wave type order parameter vector in $Sr_2RuO_4$ is contradictory to the fact that the p-wave order parameter failed to account for the data.

In this comment, contrary to Xia et al., it is demonstrated that neither the p-wave type order parameter vector, nor the asymmetry of electron-hole at the Fermi level is needed to account quantitatively for the data [1]. A magnetic field H is to make the time reversal symmetry of the system be broken. A straightforward calculation with the isotropic order parameter $\Delta(T)$, neglecting the H dependence of $\Delta(T)$ and for the pair cyclotron frequency $\Omega$ = eH/mc = (2e/2m)H/c less than the photon frequency $2\pi c/\lambda$ , yields the Kerr angle as

$\theta_K(T) = \theta_K(0)[\Delta(T)/\Delta(0)]\tanh[\Delta(T)/2k_BT]$,

where $\theta_K(0) = A \lambda^3 \Omega /(8 \pi^3 c N \lambda_L^2)$, A = $3\lambda/4\xi_{BCS}$ in the non-local (Pippard) limit and A = $L/\xi_{BCS}$ in the local (London) limit, respectively. The BCS coherence length $\xi_{BCS} = \hbar v_F/\pi\Delta(0)$. In Ref [1], the wave length $\lambda$ = 1550 nm, N = (n−1) n (n+ 1) = 24 with index of refraction n = 3, mean free path length L = 1 μm, Fermi velocity $v_F$ = 100Km/s and in Ref [2], the London penetration depth length $\lambda_L$ = 3 μm, and transition temperature $T_C$ = 1.5 K are given. In the strong coupling case [3], $\Delta(0) = 2T_C$ , which is used for our calculation. The effective magnetic field H is considered to be sum of the external applied and internal (via super-current) magnetic fields, to maintain the fluxoid quantization. After cooling a sample in the external magnetic field, turning it off, before warming a sample, is not necessary to make H vanish, since the super-current was set in a sample during cooling it. Then, $H_{C2}$ = 750 Gauss [2], in the normal vortex core, may be considered as H. Inserting all values of parameters given above in the Kerr angle formula, we obtain $\theta_K(0)$ = 44 nrad in the non-local limit, and 38 nrad in the local limit, respectively. In spite of a simple calculation, neglecting the retardation effect and others, the agreement between the calculation and measurement is satisfactory. The fluxoid quantization makes the Kerr angle same within a range of the external applied magnetic fields, as data [1] indicated. The temperature dependence of the Kerr angle is same as that of the optical super-electron density [4] as one expected.

Magnetic domains in a superconductor have been suggested some time ago [5]. The key points are; a. To have a finite transition temperature $T_C$, a finite pairing interaction energy range $T_D$ is required. The particles (states) outside of $T_D$ do not participate in pairings (Incomplete Condensation) [3], and account for the T- power behaviors of specific heat and magnetic penetration depth length at low temperature T [6], and $T_C$ reduction by the ordinary scatterings [7]. b. Consequently, the carriers not participating in pairings act as vortices, resulting in the multi-connected superconductors (MS) [8]. The notion of MS can account for data of the phase sensitive experiments such as the Josephson edge junctions, and half-flux quantum observed as well. c. The current circulating around the vortex center is found [5] to be proportional to the distance from the vortex center times the Gaussian factor, and has a maximum value at the distance $r_m = 1.0916\, \lambda_L/\kappa^{1/3}$, for a large $\kappa = \lambda_L/\xi$. This has suggested superconducting ringlets; (VF, VZF, VAF) vortex (ringlet) with fluxoid quantum (1, 0, -1), respectively, just like spin 1 states. VZF has been suggested to act as the domain wall between VF and VAF domains (regions). Now the circulating current may be represented by an equivalent torus whose center touching the vortex center. In other words, no middle empty part is allowed in the doughnut shape. Then, the radius $r_T$ of the equivalent torus (ringlet) may be determined by setting the total circulating currents in two cases be same, and has been found to be, $r_T = 0.8036\, \lambda_L/\kappa^{1/3}$, for a large $\kappa = \lambda_L/\xi$. For $\lambda_L = 3$ μm and $\kappa = 46$ [2], we get $r_T = 0.673$ μm. Then, the size of a torus (ringlet) is $4r_T = 2.7$ μm. The diameter of the smallest domain having one VF or VAF surrounded by 6 VZF, would be 5.4 μm, taking into account the half-size of a ringlet in both edges. The size of domain having (7, 19) ringlets surrounded by (12, 18) VZF would be (10.8, 16.2) μm, and so on.

The ringlet domains are just like patterns of three different colored coins laid on a table. Equivalently, they are magnetic domains in the triangular lattice with spin 1. The occurrences of VF or VAF domains are to be determined by the local conditions which are remained to be investigated. They may be responsible for the sign of the Kerr angle. Of course, the Kerr angle changes its sign at the pair cyclotron resonance frequency.

Flux bunching: flux lines are to be confined inside a ringlet, that is, they are belted in it.

It is shown that the isotropic order parameter accounts quantitatively and accurately for the data [1].

Sang Boo Nam, sangboonam@mailaps.org
7735 Peters Pike, Dayton, OH 45414-1713 USA